\documentclass[preprints,article,accept,pdftex,moreauthors]{Definitions/mdpi} 
\DeclareGraphicsExtensions{.pdf,.prb,.jpg}
\usepackage[percent]{overpic}

\usepackage[version=4]{mhchem}

\newcommand{\Lfn}[1]{{\cal L}^{(#1)}}

\newcommand{\JPB}[1]{{#1}}

\graphicspath{{./Figures_final}}

\newcommand\ve{\varepsilon}

\firstpage{1} 
\pubvolume{1}
\issuenum{1}
\articlenumber{0}
\pubyear{2025}
\copyrightyear{2025}
\datereceived{ } 
\daterevised{ } 
\dateaccepted{ } 
\datepublished{ } 
\hreflink{https://doi.org/} 

\Title{Thermoelectric Enhancement of Series-Connected Cross-Conjugated Molecular Junctions}

\TitleCitation{Thermoelectric Enhancement of Series-Connected Cross-Conjugated Molecular Junctions}

\Author{{Justin} 
 P. Bergfield \orcidA{}} 

\AuthorNames{Justin P. Bergfield}
\AuthorCitation{Bergfield, J.P.}

\address[1]{
Department of Physics, Illinois State University, Normal, IL 61761, USA; jpbergf@ilstu.edu}

\abstract{We investigate the thermoelectric response of single-molecule junctions composed of acyclic cross-conjugated molecules, including dendralene analogues and related iso-poly(diacetylene) (iso-PDA) motifs, in which node-possessing repeat units are connected in series. Using many-body quantum transport theory, we show that increasing the number of repeat units leaves the fundamental gap essentially unchanged while giving rise to a split-node spectrum whose cumulative broadening dramatically enhances the thermopower.
This form of quantum enhancement can exceed other interference-based mechanisms, such as the coalescence of nodes into a supernode, 
\textls[-15]{suggesting new opportunities for scalable quantum-interference–based materials. Although illustrated here with cross-conjugated systems, the underlying principles apply broadly to series-connected architectures hosting multiple interference nodes. 
Finally, we evaluate the scaling of the electronic figure of merit ZT and the maximum thermodynamic efficiency. Together, these results highlight the potential for split-node-based materials to realize quantum-enhanced thermoelectric~response.}}


\begin{document}

\section{Introduction}

Thermoelectric devices directly convert heat into electrical energy, making them attractive for a wide range of clean-energy applications \cite{bell2008cooling}.
At their core, these systems harness the coupling between heat and charge transport. In response to an applied temperature gradient, carriers tend to drift from hot to cold, as dictated by the {second} 
 law 
of thermodynamics, carrying both energy and entropy until an opposing potential builds and equilibrium is reached. The proportionality between the induced voltage and the applied temperature difference defines the thermopower (Seebeck coefficient). Beyond its technological importance, often quantified by the figure of merit {$ZT$,} 
 the thermoelectric response measures the interplay between entropic and electronic degrees of freedom, making it a unique probe into the underlying physics of a system \cite{evers2020advances,dubi2011colloquium,bergfield2013forty,wang2020thermal,bergfield2024identifying}.


In junctions with dimensions less than or commensurate with the deBroglie wavelength of the charge carriers, coherent wave effects can dominate transport. Here, we focus on single-molecule junctions (SMJs), open quantum systems composed of a small organic molecule coupled to macroscopic electrodes. Electronic transport through SMJs remains predominantly coherent and elastic even at room temperature and in noisy chemical or electromagnetic environments \cite{aradhya2012dissecting,arroyo2013signatures,guedon2012observation,wang2020thermal}, owing to the large charging energies of small molecules relative to the thermal energy scale. Quantum wave effects, most notably quantum interference (QI), have been directly observed in transport through SMJs \cite{guedon2012observation,li2019gate,liu2018quantum}, establishing these systems as robust testbeds for exploring coherent quantum contributions to thermoelectric and thermodynamic responses \cite{xu2022scaling}.

Of particular interest are transmission nodes, destructive QI features that occur only when all amplitudes cancel exactly. Their existence reflects underlying symmetries of the many-body Hamiltonian \cite{bergfield2011novel,barr2013transmission,pedersen2014quantum}, making them powerful indicators of a system’s fundamental structure. While certain nodes can be rationalized using topological arguments or effective single-particle models \cite{markussen2010relation,solomon2008understanding,chen2018designing,liu2018quantum}, many-body effects can qualitatively alter this picture, and simplified descriptions often fail to capture them \cite{bergfield2011novel,barr2013transmission,pedersen2014quantum}.

Transmission nodes also reveal how QI affects the thermodynamics of transport: as a node is approached, charge and entropy currents diminish at different rates, so their ratio, the entropy per unit charge (i.e., the thermopower), can be strongly enhanced \cite{bergfield2009thermoelectric,bergfield2010giant}. In this regime, the free-electron picture fails, and marked violations of the Wiedemann–Franz law emerge \cite{bergfield2009thermoelectric,bergfield2010giant,majidi2021heat}, underscoring the intrinsically quantum nature of transport. Beyond their fundamental interest, these interference phenomena demonstrate that molecular architectures can be designed to harness QI, offering a pathway toward scalable and robust thermoelectric energy-conversion materials \cite{bergfield2010giant,bennett2024quantum,bergfield2009thermoelectric,park2019structure,rincon2016thermopower,reddy2007thermoelectricity,guo2013single,guo2011measurement,evangeli2013engineering}.

Here, we focus on molecules composed of $N$ node-possessing subunits connected in series. The {\em {order} 
} of a node is defined by the leading power with which the transmission vanishes near its energy; for example, an $n${th}-order 
node satisfies ${\cal T}(\mu) \propto (\mu-\mu_{\rm node})^{2n}$, where $\mu_{\rm node}$ is the nodal energy. The case $n=1$ corresponds to a quadratic node, arising from the interference of two transmission pathways, the minimum requirement for destructive QI. 
Higher-order nodes arise when multiple nodes coincide at the same energy: when the nodal energy is independent of molecular length, the $N$ subunits combine to form a $2N$-order {\em {supernode}}, 
a feature predicted to imbue certain materials with scalable, broad-spectrum thermoelectric response \cite{bergfield2010giant}. Generally, however, the realization of supernodes appears to require finely tuned molecular parameters \cite{barr2013transmission,pedersen2014quantum}, and the generic outcome is that the spectrum exhibits $N$ distinct nodes. \JPB{The split-node regime is therefore the more general scenario, and it is the case we investigate in this work.} As we show, when the nodal density is sufficiently large, their collective influence mimics a higher-order response, effectively endowing split-node junctions with many of the thermoelectric advantages previously associated with supernodes.


To investigate these effects, we employ a many-body quantum transport theory \cite{bergfield2009many}, examining charge and heat transport through acyclic, alkynyl-extended dendralene molecules and iso-poly(diacetylene) (iso-PDA) analogues \cite{solomon2011small}. Although the strict Phelan--Orchin definition~\cite{phelan1968cross} of cross-conjugation is not essential for our purposes, these backbones contain precisely the ingredients needed to generate robust destructive interference near the electrode Fermi energy. They therefore serve as minimal interferometers, with structural parameters that allow systematic control of interference features.  

Within each cross-conjugated repeat unit, substitution at the termini of the ``stubs'' primarily renormalizes the local on-site energy without altering the connectivity responsible for interference. 
As a result, the key qualitative features, node splitting, the growth of nodal density with molecular length, and the associated thermopower response, are topologically robust to a wide range of end-group modifications. In practice, stabilizing substituents (e.g., phenyl rings, alkynyl or aryl caps, heteroatom auxiliaries) may be appended to improve kinetic stability; these mainly shift the absolute nodal energies without erasing the interference motif. The main caveat is quantitative: heavy or strongly conjugating substituents can increase $\sigma$-channel transport or broaden molecular levels, modestly reducing the thermopower amplitude, though the interference pattern itself remains intact.

Our analysis shows that while the electronic gap of these junctions remains essentially fixed, the number of interference nodes increases linearly with $N$. This scaling suggests a design principle: by tuning nodal density, one can realize enhanced thermoelectric response in specific regimes. 
Here, we examine how the thermopower and maximum thermodynamic efficiency evolve with increasing $N$, and assess the role of residual $\sigma$-channel transport.
Although we illustrate these ideas with dendralene analogues, the same principles apply to any series-connected architecture in which end-group substituents act as energy-independent onsite shifts and do not introduce additional channels. Under these conditions, nodal-density scaling and the associated thermopower enhancement emerge as generic consequences of coherent series connectivity.


\section{Quantum Transport Theory}

We investigate quantum transport through single-molecule junctions (SMJs) composed of two macroscopic electrodes, modeled as ideal Fermi gases characterized by chemical potentials and temperatures, coupled to a small organic molecule. Transport through these systems is predominantly elastic and coherent. In the linear-response regime, the key thermoelectric quantities may be expressed in terms of the Onsager {functions} 
 ${\cal L}^{(n)}$:  
\begin{align}
	G &= e^2 {\cal L}^{(0)}, \\
	S &= -\frac{1}{eT_0} \frac{{\cal L}^{(1)}}{{\cal L}^{(0)}}, \\
	\kappa_{\rm e} &= \frac{1}{T_0} \left( \Lfn{2} - \frac{\left[\Lfn{1}\right]^2}{\Lfn{0}} \right), 
	\label{eq:linear_response_variables}
\end{align}
where $e$ is the charge of the electron, $G$ is the electrical conductance, $S$ is the Seebeck coefficient, $T_0$ is the temperature, and $\kappa_{\rm e}$ is the electronic thermal conductance.  The thermoelectric device performance is often quantified by the dimensionless figure of merit \cite{bell2008cooling},
\begin{equation}
	ZT = \frac{S^2 G T_0}{\kappa_{\rm e} + \kappa_{\rm p}},
\end{equation}
where $\kappa_{\rm p}$ is the phonon contribution to the thermal conductance. Owing to the mismatch between electrode Debye frequencies and molecular vibrational modes, we neglect phonon contributions here and focus on the electronic component, denoted $ZT_{\rm e}$. 

From a thermodynamic perspective, performance is characterized by the efficiency $\eta$, defined as the ratio of useful work to input heat. While $\eta$ depends in detail on molecular structure, level alignments, and applied biases \cite{bergfield2010giant}, the maximum efficiency in linear response can be expressed in terms of $ZT_{\rm e}$ as
\begin{equation}\label{eq:eta_max}
	\frac{\eta_{\rm max}}{\eta_C} = \frac{\sqrt{1+ZT_{\rm e}} - 1}{\sqrt{1+ZT_{\rm e}} + 1},
\end{equation}
where $\eta_C$ is the Carnot efficiency \cite{thermoelements1957af}.

At room temperature, SMJ transport is primarily quantum-coherent and elastic, allowing the Onsager functions to be expressed as 
\begin{equation}\label{eq:Lfun}
{\cal L}^{(\nu)}(\mu) = \frac{1}{h} \int dE \left( E - \mu \right)^\nu {\cal T}(E) \left(-\frac{\partial f_0}{\partial E} \right),
\end{equation}
where $f_0(E) = \left[\exp((E-\mu_0)/kT_0) + 1 \right]^{-1}$ is the Fermi--Dirac distribution with chemical potential $\mu_0$ and temperature $T_0$.
We utilize nonequilibrium Green's function (NEGF) theory~\cite{HaugAndJauhoBook, stefanucci2013nonequilibrium} to describe the transport, where the transmission may be expressed in this regime as~\cite{bergfield2009many}
\begin{equation}
{\cal T}(E) = {\rm Tr} \left\{ \Gamma_L(E) {\cal G}(E) \Gamma_R(E) {\cal G}^\dag(E) \right\},
\label{eq:transmission_prob}
\end{equation}
where ${\cal G}$ 
is the junction's retarded 
Green's function. The tunneling-width matrix for contact $\alpha$ may be expressed as:
\begin{equation}
\left[\Gamma_\alpha(E)\right]_{nm} = 2\pi \sum_{k\in\alpha} V_{nk} V_{mk}^\ast\, \delta(E-\epsilon_k),
\end{equation}
where $n$ and $m$ label the $\pi$-orbitals within the molecule, and $V_{nk}$ is the coupling matrix element between orbital $n$ of the molecule and a single-particle energy eigenstate $\epsilon_k$ in electrode $\alpha$. We consider transport in the broad-band limit, treating this matrix as energy-independent.

In many-body molecular Dyson equation (MDE) theory, the Green's function of an SMJ is given by \cite{bergfield2009many}
\begin{equation}
{\cal G}(E) = \left[ {\cal G}_{\rm mol}^{-1}(E) - \Sigma_{\rm T}(E) - \Delta \Sigma_{\rm C}(E) \right]^{-1},
\label{eq:Dyson2}
\end{equation}
where ${\cal G}_{\rm mol}$ is the molecular Green's function, $\Sigma_{\rm T}$ represents the tunneling self-energy matrix with $\Sigma_{\rm T} = -i/2 \sum_\alpha \Gamma_\alpha$, and $\Delta \Sigma_{\rm C}$ is the Coulomb correction term \cite{bergfield2009many}. We restrict our attention to the elastic cotunneling regime, where $\Delta\Sigma_{\rm C}\approx0$ and inelastic processes can be neglected

The molecular Green's function, ${\cal G}_{\rm mol}$, is determined by exactly diagonalizing the molecular Hamiltonian projected onto relevant atomic orbitals \cite{bergfield2009many}:
\begin{equation}
{\cal G}_{\rm mol}(E) = \sum_{\Psi, \Psi'} \frac{[{\cal P}(\Psi) + {\cal P}(\Psi')] C^{\Psi \rightarrow \Psi'}}{E-E_{\Psi'}+E_{\Psi}+i0^+},
\label{eq:Gmol}
\end{equation}
where $E_\Psi$ is the eigenenergy of the (many-body) molecular Hamiltonian $H_{\rm mol}$, ${\cal P}(\Psi)$ is the occupation probability of state $\Psi$, and $C^{\Psi\rightarrow\Psi'}$ represents the many-body transition matrix:
\begin{equation}
C^{\Psi \rightarrow \Psi'}_{n\sigma,m\sigma'} = \langle \Psi | d_{n\sigma} | \Psi' \rangle \langle \Psi' | d^\dagger_{m\sigma'} | \Psi \rangle,
\label{eq:manybody_element}
\end{equation}
where $d_{n\sigma}$ annihilates an electron of spin $\sigma$ on the $n$th atomic orbital. Here, $\Psi$ and $\Psi'$ are eigenstates for $N$ and $N+1$ particle systems, respectively, with ${\cal P}(\nu)$ given by the grand canonical ensemble in linear response.

Although MDE is formally exact, dynamic multi-particle effects enter through the correlation to the Coulomb self-energy $\Delta\Sigma_{\rm C}(E)$, which in practice requires approximation~\cite{bergfield2009many}.
Nevertheless, a key strength of MDE theory is that it provides a nonperturbative, nonequilibrium quantum transport framework in which intramolecular correlations are treated exactly; there is no simple Wick's theorem for ${\cal G}_{\rm mol}$ (MDE is not a $\Phi$-derivable theory).
As a result, Coulomb blockade and coherent tunneling processes are treated on an equal footing, the fluctuation–dissipation theorem is satisfied, and transport in both the sequential- and cotunneling regimes is described accurately.  


\subsection{Lanczos Method for Green’s Functions}

While the Dyson equation given by Equation (\ref{eq:Dyson2}) is a formally exact expression for the Green's function, its practical evaluation in correlated systems is limited by the exponential growth of the Hilbert space. In our simulations, we include all charge and excited states of the molecule, so for $n$ orbitals, the Hilbert space contains at least $4^n$ many-body states. A powerful alternative is to employ Krylov-space techniques, e.g., Lanczos recursion~\cite{lanczos1950iteration}, to evaluate the retarded Green’s function directly from the many-body Hamiltonian. Lanczos replaces explicit diagonalization with iterative projection into a Krylov subspace, requiring only repeated applications of the Hamiltonian to a seed vector. This approach drastically reduces memory demands and converges rapidly for ground and low-lying excited states, while higher excitations are only approximated. 
\JPB{Although each Lanczos step scales only polynomially, the underlying vector dimension remains $4^n$, so the overall cost is still exponential, albeit with far smaller prefactors than matrix inversion. In practice, this enables us to reach systems up to $n \approx 16$ orbitals (e.g., the $N=3$ NCCA junction).} 
For Green’s functions, the recursion naturally generates a continued-fraction expansion of ${\cal G}$, allowing correlation functions to be evaluated without explicit knowledge of the full spectrum.


Using the method developed in ref.~\cite{barr2013transport}, the Green's function is rearranged into the~form 
\begin{equation}
\left[{\cal G}(E)\right]_{nm} = \sum_{\nu,\nu'} [P(\nu) + P(\nu')] 
\langle \nu | d^\dagger_n 
\left( \frac{|\nu'\rangle \langle \nu'|}{E - (E_{\nu'} - E_\nu) + i0^+} \right) 
d_m | \nu \rangle ,
\end{equation}
which highlights how the operators $d_m$ and $d^\dagger_n$ connect the many-body eigenstates $\{|\nu\rangle\}$ to the virtual states $\{|\nu'\rangle\}$ appearing in the propagator.
The practical implementation proceeds in two stages. First, a Lanczos diagonalization of the molecular Hamiltonian 
yields a set of low-lying eigenstates $|\nu\rangle$ and their energies $E_\nu$. Second, the action of $d_m$ or $d^\dagger_n$ on these states is used to seed a new Lanczos recursion, which generates the Krylov subspaces spanned by the virtual excitations $|\nu'\rangle$. The matrix elements and energies from this recursion enter directly into the evaluation of ${\cal G}_{nm}(E)$. In practice, one may either sum the resulting discrete contributions or employ a continued-fraction expansion \cite{mori1965continued} 
that converges rapidly and preserves causality.

This two-stage Lanczos procedure has several important advantages. It reduces the evaluation of Green’s functions in strongly correlated systems to a sequence of sparse matrix-vector multiplications, making it scalable to larger $\pi$-conjugated molecules. It also preserves the full many-body structure of the excitation spectrum, ensuring that interference effects, Coulomb blockade resonances, and nodal features are treated on an equal footing. For our purposes, the accuracy of ${\cal G}(E)$ in the vicinity of transmission nodes is essential, as these features dominate the thermopower and efficiency trends analyzed in the following sections.

\subsection{Molecular Hamiltonian}

We are interested in the response of cross-conjugated polymers, whose transport is carried predominantly by the $\pi$-system. The effective Hamiltonian for this $\pi$-subspace was derived from first principles using a renormalization procedure that incorporates off-resonant degrees of freedom (e.g., the $\sigma$-system, image-charge effects, etc.) implicitly as renormalized onsite energies and coupling terms \cite{barr2012effective}. In a basis of localized orbitals, the Hamiltonian is written  
\begin{align}
\label{eq:Hmol}
H_{\rm mol} & = \sum_{n,\sigma} \ve_{n\sigma} \hat{\rho}_{n\sigma}
- \sum_{\langle n,m \rangle,\sigma} t_{nm} \hat{d}^\dagger_{n\sigma} \hat{d}_{m\sigma} \nonumber 
  + \tfrac{1}{2} \sum_{nm} U_{nm} \hat{q}_n \hat{q}_m ,
\end{align}
where $\ve_{n\sigma}$ is the effective onsite potential for spin-$\sigma$ electrons on orbital $n$, 
$\hat{\rho}_{n\sigma} = \hat{d}^\dagger_{n\sigma} \hat{d}_{n\sigma}$, 
$\hat{q}_{n} = (\sum_\sigma \hat{\rho}_{n\sigma})-1$ is the net charge operator, and $t_{nm}$ are the effective tight-binding matrix elements. The Coulomb interaction $U_{nm}$ between electrons in orbitals $n$ and $m$ is obtained from a multipole expansion including monopole--monopole, quadrupole–monopole, and quadrupole--quadrupole terms \cite{barr2012effective}:  
\begin{align}
U_{nm} &= U_{nn}\,\delta_{nm} + (1-\delta_{nm})\!\left(U^{MM}_{nm} + U^{QM}_{nm} + U^{MQ}_{nm} + U^{QQ}_{nm}\right).
\end{align}
%

The \JPB{$\pi$-system's effective field theory (EFT)} parameters were obtained through a renormalization procedure in which experimental observables were fit to quantities that must be faithfully reproduced by a $\pi$-electron-only model \cite{barr2012effective}. Specifically, the vertical ionization energy, the vertical electron affinity, and the six lowest singlet and triplet excitations of neutral gas-phase benzene were simultaneously optimized \cite{barr2012effective}. This procedure yields a fit that is comparable to, or better than, traditional  Pariser--Parr--Pople (PPP) models \cite{barford2005electronic}, yielding $U_{nn}=9.69$~eV for the onsite repulsion, transfer integrals $t=2.2$, $2.7$, and $3.0$ eV for single, double, and triple carbon--carbon bonds, respectively, and a $\pi$-electron quadrupole moment $Q = -0.65$ e\AA$^2$. Interactions are screened by a uniform dielectric constant 
$\varepsilon=1.56$. 
These values are consistent with earlier $\pi$-electron models \cite{barford2005electronic,ohno1964some}, with $Q$ providing a physically motivated alternative to the ad hoc short-range corrections of PPP theory.

The electrodes were modeled as metallic spheres of radius 0.5 nm. The partial ionic character of the Au--S bond was represented by point charges of $-$0.67{e} 
 placed at the sulfur positions \cite{barr2012effective}, determined by a simultaneous fit to experimental thermopower~\cite{baheti2008probing} and conductance~\cite{xiao2004measurement} of benzene-based junctions \cite{bennett2024quantum,bergfield2011number,bergfield2012transmission}. The screening surface was taken one covalent radius beyond the outermost Au nucleus \cite{PhysRevB.30.5669,cordero2008covalent}.
\JPB{Molecular geometries were optimized with Kohn--Sham DFT in ORCA 6.1.0 at the B3LYP-D3(BJ)/6-311G(d,p) level (6-311G**), using tight SCF thresholds; harmonic frequency analyses at the same level confirmed all stationary points as minima (no imaginary modes) \cite{Becke1993,LYP1988,Krishnan1980,McLeanChandler1980,  RN269, RN33, RN75,RN114,RN218}}.
Image-charge effects were incorporated through the renormalization procedure.

\section{Thermoelectric Response}
\begin{figure}[htb]
	\includegraphics[width=0.95\linewidth]{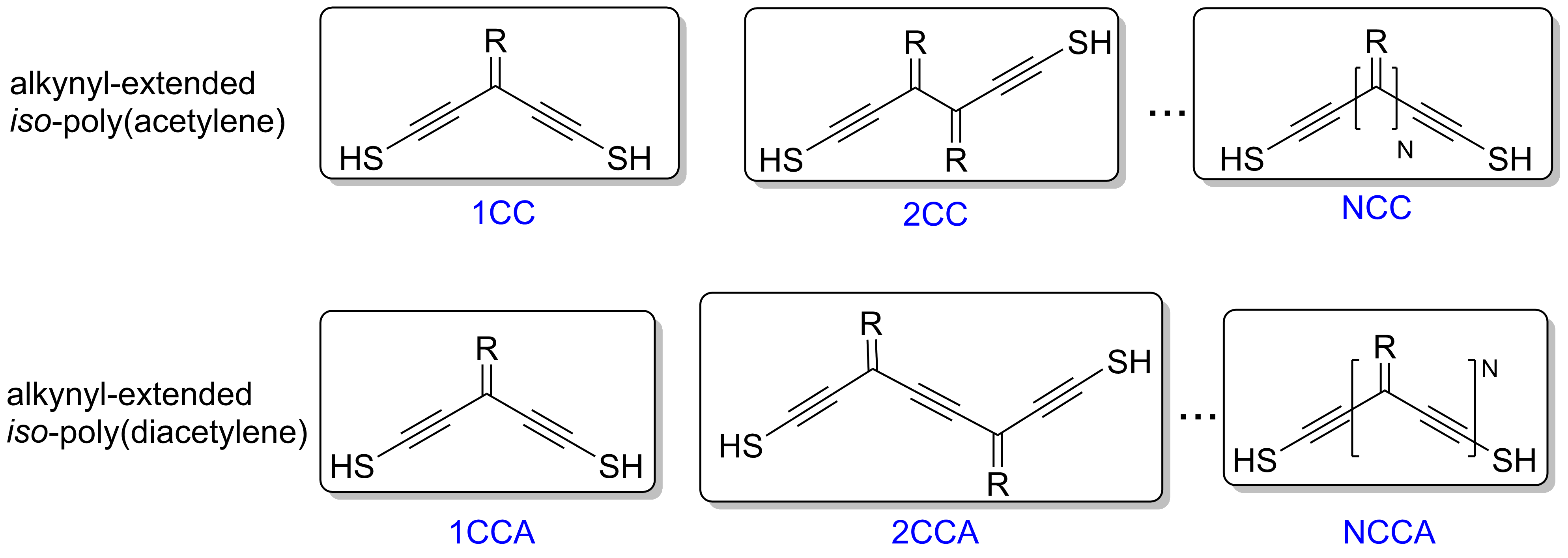}
    \caption{{Schematic} 
 diagrams of thiolated, alkynyl-extended iso-poly(acetylene) (\textbf{top}) and iso-poly(diacetylene) (\textbf{bottom}) analogues, denoted NCC and NCCA, respectively. Each cross-conjugated repeat unit contains terminal ``stub'' {C--C} groups with substituents $R$. In principle, these may be varied to stabilize the backbone or tune onsite energies. In this work, we set $R=H$ to isolate the intrinsic interference features.
    }
	\label{fig:molecular_motif}
\end{figure}

We investigate the thermoelectric response of SMJs composed of molecules built from repeated node-possessing subunits. Because the many-body Hilbert space grows exponentially with system size, we begin with the minimal node-bearing motif: the cross-conjugated diene (a single \ce{C=C} stub unit) \cite{solomon2011small}, which extends to form the dendralene family. Dendralenes represent the simplest class of cross-conjugated oligoenes and provide a natural starting point for exploring interference-driven thermoelectric response.  

Our focus is on alkynyl-extended dendralene analogues (henceforth, the NCC series, with \ce{R=H}), shown in Figure~\ref{fig:molecular_motif}, bonded to Au electrodes via thiol linkers. Alternative stabilizing substituents \ce{R} may be used in practice, provided they do not introduce strong energy dependence or significant coupling to additional degrees of freedom. Otherwise, such groups act as dephasing probes and diminish the coherent interference response \mbox{\cite{bergfield2025quantuminterferencesupernodesthermoelectric,erdogan2025dephasingfailsthermodynamicconsequences}.} To leading order, substituents that respect these constraints serve only to renormalize the onsite energies and do not substantively alter the findings presented here. 

\begin{figure}[tb] 
	\centering 
	\includegraphics[width=0.95\linewidth]{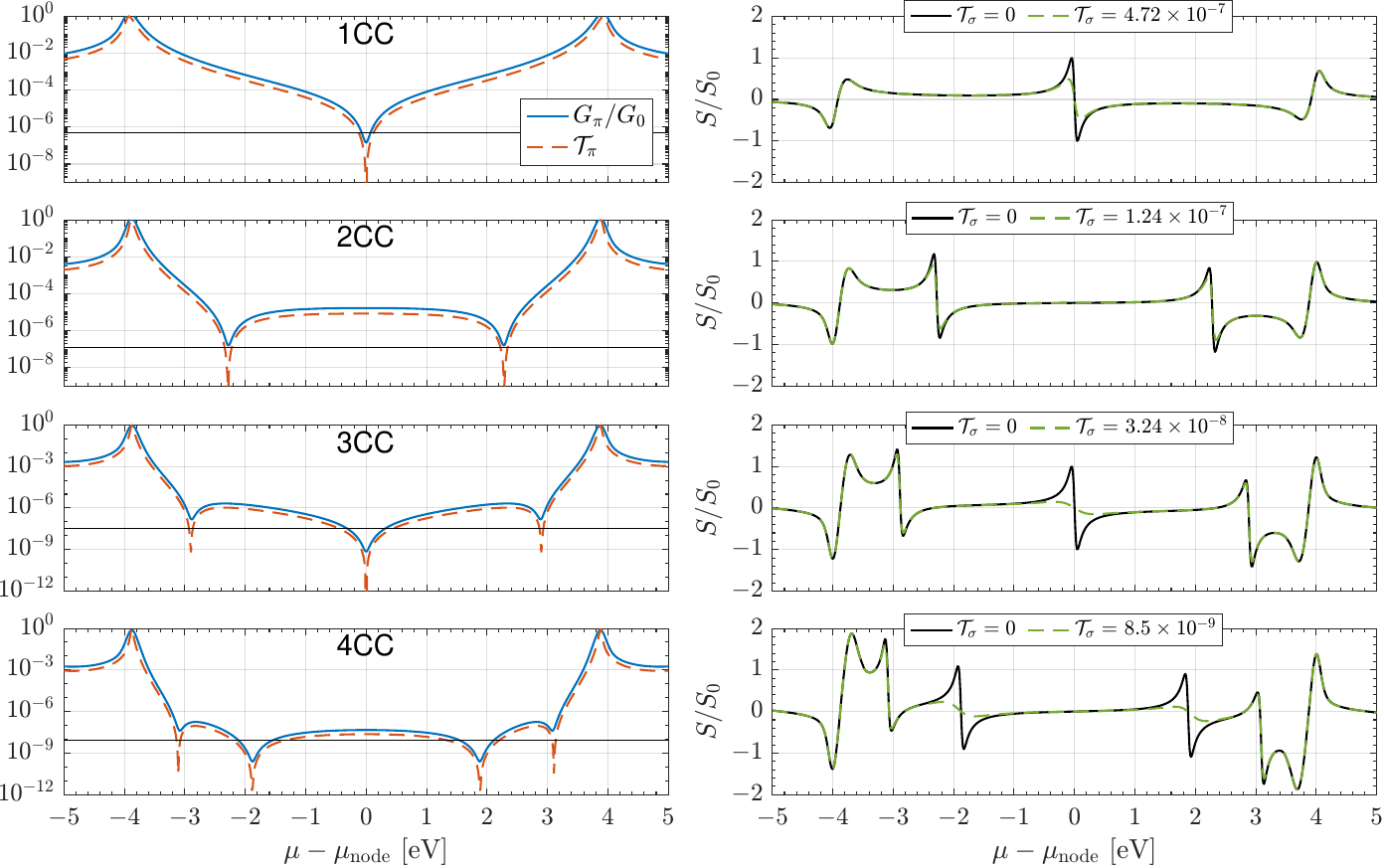}
 \caption{{Many-body} 
 calculated transmission probability $\mathcal{T}_\pi$ and
    conductance $G_\pi$ (\textbf{left} panels) and total thermopower $S$ (\textbf{right} panels) for NCC
    junctions with $N$ = 1--4, shown as a function of electrode chemical
    potential $\mu$. The mid-gap energy is set to zero in all cases. Each
    repeated unit introduces one quadratic transmission node, so the total
    number of nodes equals $N$, while the fundamental gap remains nearly
    independent of $N$. 
    \JPB{The HOMO and LUMO features near $\pm 4$ eV correspond to the many-body addition/removal resonances ($N-1 \to N$ and $N \to N+1$ charging excitations) of the $\pi$-system, rather than Kohn–Sham orbital levels. The nearly constant gap reflects the repeat-unit conjugation length, not the overall molecular size.} Black lines in the left panels show the
    $\sigma$-channel background ($\mathcal{T}_\sigma$ with $\beta_\sigma =$ 1 \AA$^{-1}$, $A_\sigma=10^{-4}$),
    included to illustrate how even a small energy-independent $\sigma$
    contribution can obscure $\pi$-system variations. The thermopower exhibits
    sharp peaks near each node, normalized to the quadratic-node value
    $S_0=\pi/\sqrt{3}\,(k_B/e) \approx 156$ {$\upmu$V/K}. 
    As $N$ increases, overlapping tails of adjacent
    peaks enhance the mid-gap response. Dashed lines show results with nonzero
    $\mathcal{T}_\sigma$, which wash out the interference-induced thermopower
    enhancement. Conductance is normalized to the quantum $G_0=2${e$^2$/h.
    } 
    Calculations correspond to room temperature, $T_0=300$~K.}
	\label{fig:NCC}
\end{figure}

The branched structure of 
these molecules can be viewed as $N$ cross-conjugated subunits connected in series. Each branch supports both direct and indirect transmission amplitudes, and it is their coherent interplay that produces the nodal structure in these junctions \cite{barr2013transmission,solomon2011small}. The alkynyl extensions act as insulating standoffs, attenuating through-bond $\sigma$ transport. This suppression of the $\sigma$ background is essential for the node-enhanced thermoelectric response of the $\pi$ system to be observed experimentally.  

The calculated $\pi$-system transmission function ${\cal T}_\pi$ and conductance $G$ through several NCC junctions are shown in the left-hand panels of Figure~\ref{fig:NCC} as a function of the electrode chemical potential $\mu$. All calculations employ MDE many-body theory, which includes all charge and excited states of the molecule. Optimized molecular geometries were held fixed during the transport calculations, and molecule–electrode hybridization was taken as symmetric, with couplings $\Gamma_L = \Gamma_R = 0.5$ eV. All spectra were shifted such that the particle--hole symmetric point lies at $\mu=0$, and results correspond to junctions operating at room temperature ($T_0=300$ K).

As seen in the top panel of Figure~\ref{fig:NCC}, the 1CC junction exhibits a single quadratic node in its low-energy spectrum. For larger $N$, the spectra are further suppressed and display multiple split nodes: $N$ distinct nodes for an $N$-unit NCC molecule, in agreement with prior predictions \cite{barr2013transmission,pedersen2014quantum,solomon2011small,bergfield2011novel}.

%
%
Owing to the molecular symmetry and repeating subunit motif, the fundamental charging energy $U$ (closely related to the HOMO–LUMO gap once excitonic and charging effects are included~\cite{bergfield2009many}) is determined primarily by the conjugation length of the repeat unit rather than by the overall molecular length~\cite{limacher2011cross}.
\JPB{This is reflected in the weak dependence of the frontier addition and removal resonances, conventionally denoted HOMO and LUMO, on $N$ in Figure~\ref{fig:NCC}. These resonances correspond to the $N-1 \to N$ and $N \to N+1$ charging excitations of the many-body $\pi$-system, not to single-particle orbital energies \cite{perdew1982density,sham1983density,onida2002electronic}}.
Consequently, increasing $N$ leaves the fundamental gap essentially unchanged but increases the nodal density, $\Delta \sim U/N$, even though the individual node spacings are not uniform.


Each transmission node is accompanied by a peak in the thermopower spectrum, as shown in the right-hand panels of Figure~\ref{fig:NCC}. The vertical axis is normalized to the peak value $S_0 = \pi/\sqrt{3} (k_B/e) \approx 156$ {$\upmu\text{V/K}$} 
 expected for a quadratic node \cite{bergfield2009thermoelectric,bergfield2010giant,bennett2024quantum}, providing a natural reference scale. 
When multiple interference nodes are present, the tails of neighboring nodes overlap, enhancing the junction's thermopower across an extended energy window. As $N$ increases, the density of nodes grows, amplifying the variation in ${\cal T}_\pi$ with energy and producing a cumulative enhancement that scales with molecular length.
\begin{figure}[tb] 
	\centering 
 	\includegraphics[width=\linewidth]{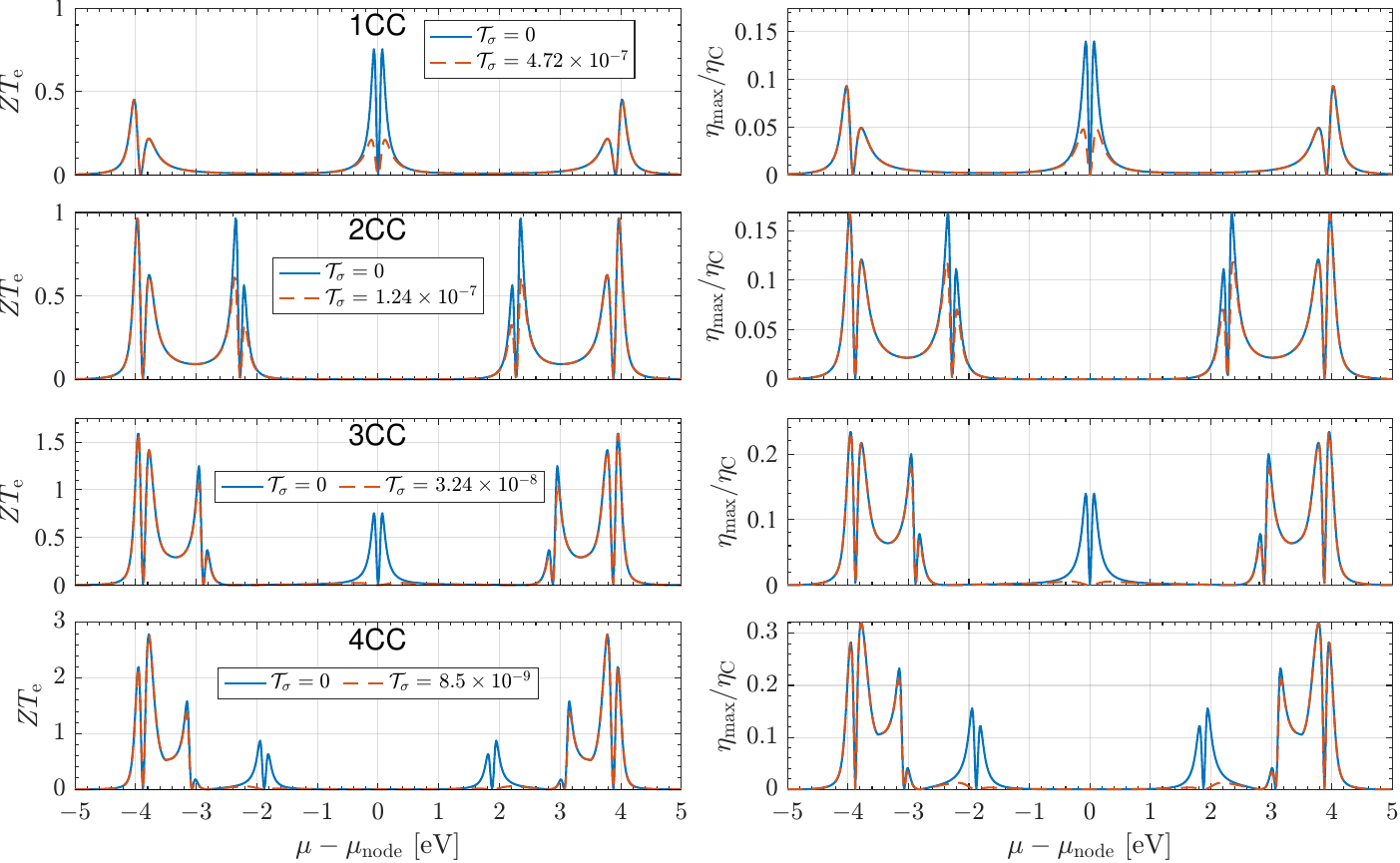}
\caption{{Many-body} 
 calculated total $ZT_{\rm e}$ (\textbf{left} panels) and total normalized maximum efficiency $\eta_{\rm max}/\eta_{\rm C}$ (\textbf{right} panels) for NCC junctions with $N$ = 1--4. In 1CC, the mid-gap node is strongly affected by $\sigma$ leakage, with peak $ZT_{\rm e}$ reduced from $\sim$0.75 to $\sim$0.21. For 2CC, the split nodes lie nearer the frontier resonances, and the suppression is lessened ($\sim$0.96 to $\sim$0.61). In 3CC, the central mid-gap node collapses ($\sim$0.75 to $\sim$0.022), but the side nodes retain large values ($\sim$1.25 to $\sim$1.06). By 4CC, additional nodes crowd the resonance regions, yielding near-resonant peaks exceeding 1.5 that remain largely insensitive to $\sigma$ transport. $\sigma$-channel transport is estimated using $\beta_\sigma=1.0$ \AA, and $A_\sigma$ = $10^{-4}$. All calculations correspond to $T_0=300$ K.
}
	\label{fig:NCC_ZT_eta}
\end{figure}

The electronic figure of merit $ZT_{\rm e}$ and the normalized maximum efficiency $\eta_{\rm max}/\eta_{\rm C}$ are shown in Figure~\ref{fig:NCC_ZT_eta} for NCC junctions with $N$ = 1--4. As with the thermopower $S$, nodes enhance both quantities \cite{bergfield2010giant}. Unlike $S$, however, they also depend on $G$ and $\kappa_e$, making them especially vulnerable to $\sigma$-channel leakage. In the $\pi$-system, peak $ZT_{\rm e}$ values exceed unity, a benchmark often cited as technologically significant \cite{bell2008cooling}, although inclusion of the $\sigma$ background reduces performance in the mid-gap region.

By contrast, interference-driven enhancements of these quantities are far more robust near the frontier resonances. For 1CC, there is only a mid-gap node, so the high-impedance regime renders $ZT$ particularly fragile: $\sigma$ leakage reduces the peak value by about 72\%. In 2CC, the split nodes lie closer to the HOMO or LUMO resonances, and the associated near-resonant enhancements are only moderately diminished, with peak values reduced by roughly 35\%. In 3CC, this trend intensifies: the central mid-gap node collapses under $\sigma$ leakage (a reduction of about 97\%), while the side nodes retain most of their strength, reduced by only $\sim$15\%. By 4CC and beyond, additional nodes crowd the frontier region, producing scalable near-resonant enhancements that remain largely insensitive to $\sigma$ leakage. These results suggest a practical strategy: tune contact chemistry and electrode work function to place $\mu$ near a near-resonant split node while minimizing $\sigma$ coupling (e.g., sp-rich linkers, top-site or tilted binding), thereby maximizing $ZT$ or $\eta_{\max}/\eta_{\rm C}$ with minimal $\sigma$ influence.

Both the thermoelectric and thermodynamic enhancement of node-possessing junctions can be understood from the Mott formula, which relates the Seebeck coefficient to the logarithmic derivative of the transmission function. In the special case where all nodes combine into a single $2N$-order supernode, the thermopower increases essentially linearly with molecular length \cite{bergfield2010giant}. The realization of such a supernode requires the nodal energy of each repeat unit to remain nearly independent of $N$, a condition that appears to depend sensitively on finely tuned molecular parameters \cite{barr2013transmission}. In practice, supernodes generally split into $N$ separate quadratic nodes \cite{barr2013transmission,pedersen2014quantum}; however, the essential idea of node-order enhancement remains: as $N$ increases with a nearly fixed gap, the nodal density, and the {\em {effective} 
} node order, increase, giving rise to additional thermopower enhancement.

This leads to the central question of the present work: can the thermoelectric response of a junction with split nodes actually surpass that of a junction hosting a supernode? To answer this, we next incorporate realistic estimates of the $\sigma$-channel contribution, which tends to wash out the variations in the $\pi$-system, and discuss strategies for reducing its impact. We then develop a low-energy analytic model that captures the nodal physics of long chains and enables direct comparison between the thermoelectric performance of split-node and supernode junctions.

\subsection{The Influence of $\sigma$-System Transport}

The total transmission through these SMJs is composed predominantly of contributions from the $\pi$ and $\sigma$ channels,  

\begin{equation}
{\cal T}_{\rm tot} = {\cal T}_\pi + {\cal T}_\sigma,
\end{equation}
where ${\cal T}_\sigma$ denotes the $\sigma$-system transmission function. In linear response, the total Onsager coefficients can be expressed as  
\begin{equation}
\Lfn{\nu}_{\rm tot} = \Lfn{\nu}_\pi + \Lfn{\nu}_\sigma.
\end{equation}
{For} 
 these systems, ${\cal T}_\sigma$ is nearly energy-independent in the mid-gap window \cite{solomon2008understanding}, giving  
\begin{equation}
\Lfn{0}_\sigma=\frac{\mathcal T_\sigma}{h},\qquad
\Lfn{1}_\sigma=0,\qquad
\Lfn{2}_\sigma=\frac{\mathcal T_\sigma}{h}\,\frac{\pi^2}{3}(k_BT_0)^2.
\end{equation}
{The}  
 observable transport coefficients then follow as  
\begin{equation}
G=e^2\Lfn{0}_{\rm tot},\qquad
S=-\frac{1}{eT_0}\frac{\Lfn{1}_{\rm tot}}{\Lfn{0}_{\rm tot}},\qquad
\kappa_e=\frac{1}{T_0}\left(\Lfn{2}_{\rm tot}-\frac{(\Lfn{1}_{\rm tot})^2}{\Lfn{0}_{\rm tot}}\right).
\label{eq:GSkappa_again}
\end{equation}
{Because} 
 $\Lfn{1}_\sigma=0$, the numerator of $S$ is unaffected by the $\sigma$ channel while the denominator is increased. The total thermopower is therefore reduced according to  
\begin{equation}
S_{\rm tot}=S_\pi\,\frac{\Lfn{0}_\pi}{\Lfn{0}_\pi+\Lfn{0}_\sigma}.
\label{eq:Stot}
\end{equation}

The same logic carries over to the figure of merit $ZT$, but here the effect is even more pronounced. Since $ZT \propto S^2$, the suppression of $S$ enters quadratically. At the same time, $\kappa_e$ is increased by the Wiedemann–Franz contribution of the $\sigma$ background together with a mixing term proportional to $S_\pi^2$. Thus, even modest $\sigma$ leakage simultaneously reduces $S$, diminishes $S^2$, and increases $\kappa_e$, yielding a disproportionately strong degradation of $ZT$ relative to the ideal $\pi$-system, as shown in Figure~\ref{fig:NCC_ZT_eta}.

Although the $\sigma$-system is formally included via the renormalized parameters of the $\pi$-EFT Hamiltonian, its magnitude can be estimated independently from benchmarks on saturated alkanes. Recent measurements have established a robust exponential law for off-resonant $\sigma$ tunneling \cite{van2022benchmark, guo2011measurement},
\begin{equation}
  \mathcal{T}_\sigma(L) = A_\sigma e^{-\beta_\sigma L},
\end{equation}
with decay constants $\beta_\sigma \simeq 1.05 \pm 0.08$ per \ce{CH2} ($\approx$$0.8$--$1.0$~\AA$^{-1}$) for thiols and $\beta_\sigma \simeq 0.8$ per \ce{CH2} ($\approx$$0.6$~\AA$^{-1}$) for amines. These values reproduce the canonical C3/C6/C8 conductance peaks ($G/G_0$$\sim$$10^{-3}$--$10^{-5}$) and serve as accepted benchmarks for the $\sigma$ channel. Earlier break-junction studies reported similar values \cite{li2006conductance,li2008charge,haiss2009impact,venkataraman2006single}, and transition-voltage spectroscopy further confirmed that $\beta_\sigma$ correlates with the tunneling barrier height \cite{guo2011measurement}. Together, these observations reinforce the view that $\sigma$ transport is largely energy-independent in the mid-gap window considered here.

If one adopts the Au--S prefactor $A_\sigma \approx 0.1$ directly, the $\sigma$-channel transmission would already be of order $10^{-4}$ at molecular spans of $L$$\sim$6--8~\AA, comparable to the measured conductances of hexanedithiol and octanedithiol. Such a contribution would overwhelm the nodal variations in $\mathcal{T}_\pi(E)$ and wash out the associated thermopower enhancement. In practice, however, the effective $A_\sigma$ depends sensitively on the contact chemistry and geometry. Weaker anchoring groups such as amines, pyridyls, or sp-hybridized alkynyl linkers, as well as top-site or tilted configurations, can reduce $A_\sigma$ by one to two orders of magnitude.  

For illustration, in the NCC calculations shown in Figure~\ref{fig:NCC}, we set $\beta_\sigma = 1$~\AA$^{-1}$ and $A_\sigma = 10^{-4}$. As indicated in the right-hand panel, these parameters yield ${\cal T}_\sigma$ values that only partially reduce the node-enhanced thermoelectric response. This robustness motivates a focus on backbones with intrinsically weaker $\sigma$ overlap. A natural, but still computationally tractable choice is iso-polydiacetylene (iso-PDA) analogues \cite{solomon2011small}, whose longer sp-rich scaffolds preserve $\pi$ conjugation while further suppressing through-bond $\sigma$ transport.

\subsection{{Iso-PDA} 
 Junctions}

\begin{figure}[H]
	\centering 
	\includegraphics[width=\linewidth]{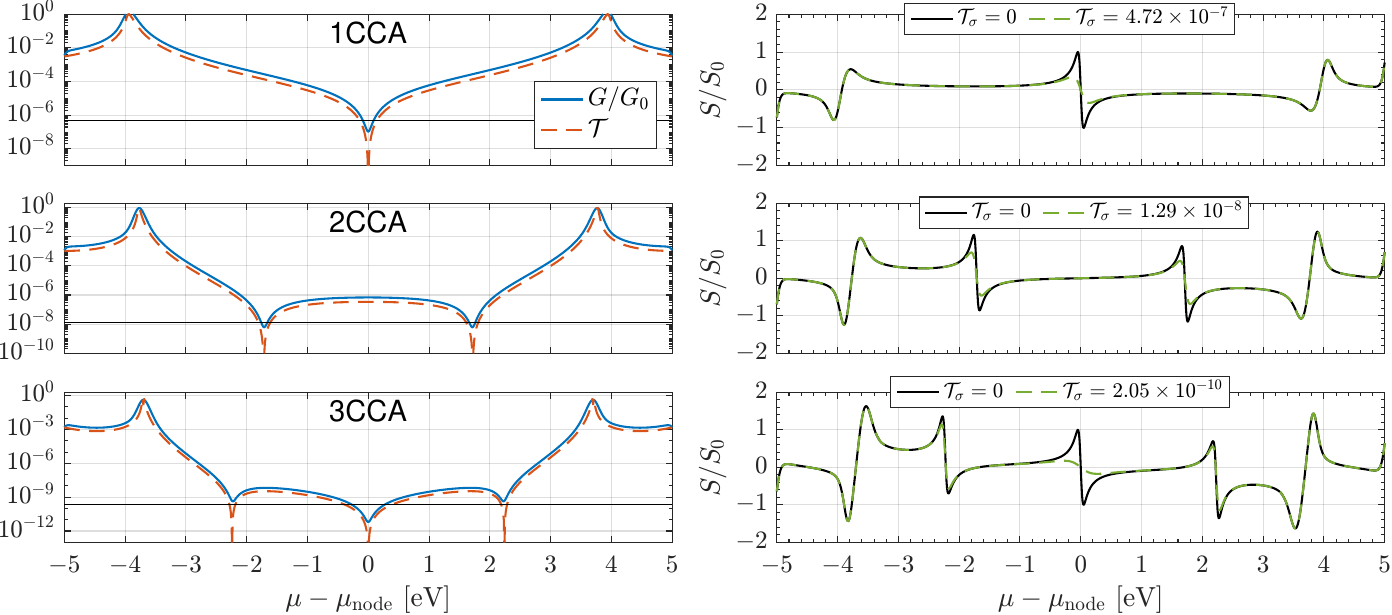}
\caption{{Many-body} 
 calculated transmission $\mathcal{T}$ and conductance $G$ (\textbf{left} panels) and thermopower $S$ (\textbf{right} panels) for NCCA junctions with $N=1$--3, shown as functions of electrode chemical potential $\mu$ (mid-gap set to zero). Each repeat unit introduces a transmission node, while the fundamental gap remains nearly independent of $N$, though slightly reduced relative to the NCC series. The longer saturated backbones of the NCCs suppress the $\sigma$-channel background, as illustrated by the black lines in the left panels ($A_\sigma=10^{-4}$, $\beta_\sigma=1.0$), thereby preserving the $\pi$-system interference pattern and allowing the nodal thermopower enhancement to emerge more clearly than in Figure~\ref{fig:NCC}. Conductance is normalized to the quantum $G_0=2${e$^2$/h},
 thermopower to the quadratic-node peak value $S_0=\pi/\sqrt{3}\,(k_B/e)$, and all calculations correspond to room temperature ($T_0=300$~K).}
	\label{fig:NCCA}
\end{figure}

Motivated by the need to suppress $\sigma$-channel transport, we consider thiolated iso-PDA analogues, denoted here as NCCA \cite{solomon2011small}, shown schematically in the lower panels of Figure~\ref{fig:molecular_motif}. The many-body calculated transmission, conductance, and thermopower for $N$ = 1--3 junctions are presented in Figure~\ref{fig:NCCA}. \JPB{The $N=3$ case already entails an active $\pi$ space of $n=16$ orbitals, making the exact many-body calculation nontrivial.} As in the NCC series, each repeat unit introduces a quadratic transmission node distributed across the mid-gap region, while the fundamental gap remains nearly independent of $N$.


For the $\sigma$-channel background, we retain the decay constant $\beta_\sigma= 1.0$~\AA$^{-1}$ and prefactor $A_\sigma = 10^{-4}$ used previously. However, owing to the additional length of the sp-hybridized diacetylene linkers, the effective influence of $\sigma$ transport is greatly reduced. As shown by the black curves in the left-hand panels of Figure~\ref{fig:NCCA}, this suppression preserves the nodal interference structure of the $\pi$-system. Quantitatively, ${\cal T}_\sigma$ is diminished by roughly one and three orders of magnitude for $N=2$ and $N=3$, respectively, relative to the NCC series. 
This robustness suggests that end-group and substituent engineering provide a practical route to realizing $\pi$-dominated thermopower enhancements.

\subsection{Low-Energy Model for Split-Node Transport}

While full many-body calculations are computationally prohibitive for large $N$, the trend is unambiguous: repeated subunits produce a split-node spectrum rather than a supernode, with a nodal density $\Delta$ that grows linearly with molecular length. To elucidate how this scaling shapes the thermoelectric response, and to identify the peak performance attainable, we introduce a simple low-energy model that encapsulates the essential transport physics of longer chains. Since transport is governed primarily by states within $k_B T_0$ of the electrode chemical potential, we approximate the $\pi$-system transmission by a sequence of equally spaced quadratic nodes,
\begin{equation}
\label{eq:Tmodel}
\mathcal{T}_\pi(E)\propto E^2 \prod_{m=1}^M (E-m\Delta)^2 (E+m\Delta)^2,
\end{equation}
where $\Delta$ denotes the node spacing and $E=0$ defines the central node energy. Supernode response is recovered when $\Delta = 0$. Apart from an overall prefactor, this form represents a molecule with $N=2M$ {+} 1 quadratic nodes and captures the essential structure needed for analytic evaluation of the Onsager functions (see Appendix \ref{appendixa}).


The optimal spacing for the thermopower, $\Delta_{\rm peak}^{S}$, is obtained by maximizing $S$ with respect to $\mu$, i.e.,
 \begin{equation}
 \left.\frac{\partial S}{\partial \mu}\right|_{\Delta_{\rm peak},\,\mu_{\rm peak}}=0.
 \end{equation} 
{For}
 $N=3$, numerical maximization gives
\begin{equation}
\Delta_{\rm peak}^{S} \approx 3.40\,k_BT_0,\qquad 
\mu_{\rm peak}^{S} \approx 1.92\,k_BT_0,
\end{equation}
in excellent agreement with the quadratic-node estimate $|\mu_{\rm peak}-\mu_{\rm node}|=\pi/\sqrt{3}\,k_BT_0$, which yields $\Delta_{\rm peak}\approx3.36\,k_BT_0$ (see Appendix \ref{appendixa}). 
A similar analysis for $ZT_{\rm e}$ (and hence $\eta_{\rm max}$) gives
\begin{equation}
\Delta_{\rm peak}^{ZT} \approx 3.52\,k_BT_0,\qquad
\mu_{\rm peak}^{ZT} \approx 3.14\,k_BT_0,
\end{equation}
so that the efficiency optimum occurs at slightly larger node spacing and deeper chemical potential than the thermopower optimum.
%
%
%
%
Numerically, we obtain $\max(|S|) \approx 501~\upmu\mathrm{V/K}$,
compared to $\approx 432~\upmu\mathrm{V/K}$ for the corresponding supernode, an enhancement of $\sim$16\%. Likewise, $\max(ZT_{\rm e}) \approx 4.0$, compared to $\approx$2.93 for the supernode \cite{bergfield2010giant}, an enhancement of $\sim$37\%. The associated maximum efficiency increases from $\eta_{\max}/\eta_C\approx 0.32$ for the supernode to $\approx$0.38 for the split-node, a gain of $\sim$19\%. A weak $\sigma$-channel background attenuates the absolute peak values of $S$ and $ZT_{\rm e}$ but leaves the optimal node spacing $\Delta_{\rm peak}$ essentially unchanged.

\begin{figure}[htb] 
\includegraphics[width=0.48\linewidth]{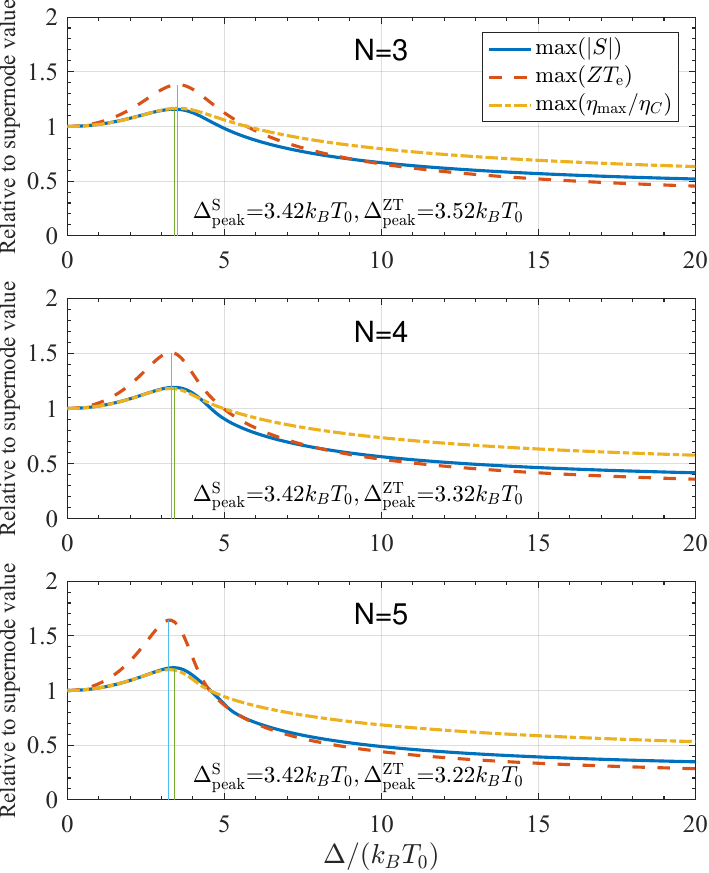}
\includegraphics[width=0.48\linewidth]{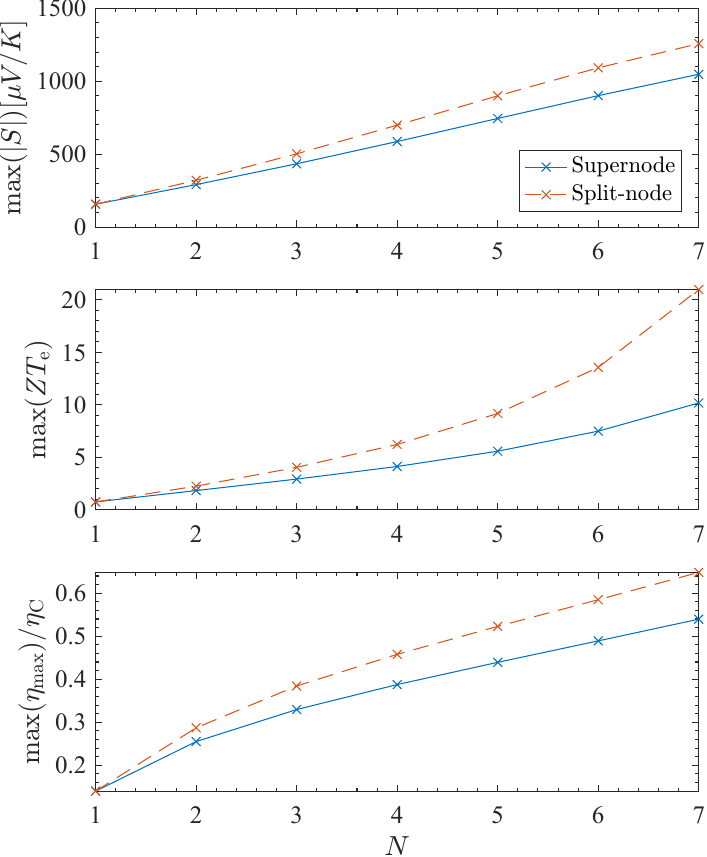}
\caption{{Maximum} 
 thermopower $|S|$, electronic figure of merit $ZT_{\rm e}$, and normalized maximum efficiency $\eta_{\max}/\eta_{\rm C}$ obtained from the split-node model of Equation~(\ref{eq:Tmodel}). The left panel shows each quantity as a function of node spacing $\Delta/(k_B T_0)$ for several chain lengths $N$, while the right panel shows the global maxima of the same quantities as a function of $N$ with ${\cal T}_\sigma=0$. In all cases, the optimum occurs at $\Delta/(k_B T_0)\simeq 3.4$--$3.5$, essentially independent of $N$, whereas the attainable maxima increase systematically with molecular length. (The $\Delta=0$ limit corresponds to the supernode case). \JPB{The green and blue vertical lines correspond to the peak splitting for $S$ and $ZT$, respectively.} The relative enhancement of split-node junctions over the supernode grows further with $N$, highlighting the parameter window where series-connected, node-bearing architectures can outperform the supernode limit.}
\label{fig:model_results}

\end{figure}

The maximum values of $|S|$, $ZT_{\rm e}$, and $\eta_{\rm max} / \eta_{\rm C}$ are plotted in the left-hand panels of Figure~\ref{fig:model_results} as a function of splitting $\Delta / k_B T_0$ for $N=3,4,5$ using the model $\pi$-system transmission. The optimal spacing for each thermodynamic quantity is essentially identical and nearly independent of $N$: additional nodes primarily increase the peak magnitude and broaden the response, while the optimal ratio $\Delta/k_B T_0$ is set by the nearest neighbors at $\pm\Delta$. 
Importantly, the advantage of split-node junctions over supernodes persists across a wide range of $\Delta$ values.


Finally, the right-hand panel of Figure~\ref{fig:model_results} shows the maximum attainable values for these quantities as a function of $N$ for both split-node and supernode junctions. While $\Delta$ is nearly $N$-independent, the maxima of all quantities grow (super)linearly with the number of subunits. This reflects the increased variation in $\mathcal{T}_\pi(E)$ in the mid-gap region, which sharpens the energy derivative of the conductance and amplifies the thermopower. Consequently, the position of the optimum is dictated by the local nodal environment, whereas the attainable magnitude is governed globally by the number of repeated units, a growth that ultimately surpasses the impressive scaling predicted for ideal supernodes~\cite{bergfield2010giant}.

\JPB{
While our investigation has focused on cross-conjugated molecules, the split-node mechanism is expected to apply to any acyclic connection of node-bearing subunits~\mbox{\cite{barr2013transmission,pedersen2014quantum,solomon2011small}} and is therefore not restricted to a particular chemistry. For example, a split-node spectrum has also been identified in multi-phenyl molecular junctions such as biphenyl \cite{barr2013transport}, suggesting that the enhancement discussed here may be more broadly realizable. Experimentally, single-molecule break-junction and STM-based thermopower measurements, together with emerging scanning thermopower imaging, provide viable routes to probe the predicted scaling with $N$ and node spacing $\Delta$. The principal challenges for such measurements are maintaining molecular stability at increasing length, achieving reproducible contacting, mitigating dephasing \cite{bergfield2025quantuminterferencesupernodesthermoelectric}, and minimizing $\sigma$-channel leakage. Nonetheless, our analysis indicates that the $\pi$-channel contribution can remain sufficiently strong for split-node signatures to be experimentally resolvable.
}

\section{Conclusions}

Using state-of-the-art many-body transport calculations, we have shown that series-connected, node-possessing molecular junctions exhibit a characteristic split-node spectrum: each repeat unit contributes an interference node, while the fundamental charging energy is set primarily by the conjugation length of the unit. As a result, the nodal density within the transport window increases with molecular length, producing strong enhancements of the thermopower and related thermodynamic quantities. This scaling sharpens the energy derivative of the transmission and yields pronounced gains in the figure of merit and efficiency.

While supernodes, $2N$-order interference features, have been predicted to imbue certain systems with significant, scalable, broad-spectrum thermoelectric response \cite{bergfield2010giant}, subsequent work has shown that such behavior requires exact symmetries or fine-tuned parameters, with splitting the more generic outcome \cite{barr2013transmission}. Our many-body calculations confirm this splitting in alkynyl-extended dendralene and iso-PDA junctions, and further demonstrate that the resulting architectures not only sustain substantial thermoelectric response over a broad energy range but can in fact outperform their supernode counterparts.

The thermoelectric enhancement in split-node systems is governed by a balance: if $\Delta$ is too small, thermal averaging smears out interference, while if $\Delta$ is too large, the influence of neighboring nodes becomes negligible. The optimal node spacing, $\Delta_{\rm peak}\simeq 3.4$--$3.5\,k_B T_0$, emerges consistently within our model, while the attainable response grows systematically with $N$. Since the fundamental gap $U$ sets the spectral width, this yields a simple design rule:
\begin{equation}
N_{\rm opt} \simeq \frac{U}{3.4\,k_B T_0}, 
\end{equation}
indicating that one may either select $N$ and tailor the repeat-unit conjugation to set $U$, or begin from a known $U$ and choose the backbone length to maximize performance. At this spacing, the maxima of $|S|$, $ZT_{\rm e}$, and $\eta_{\rm max}/\eta_{\rm C}$ exceed those of the corresponding supernode, with the relative advantage increasing systematically with molecular length.

Suppressing $\sigma$-channel leakage remains essential for translating favorable $\pi$-system thermopower into practical efficiency, but iso-PDA scaffolds and other sp-rich linkers provide promising structural routes. As with supernode-based thermoelectric materials, split-node architectures are most naturally suited to high-impedance applications. Moreover, because thermopower is an extensive quantity, the node-enhanced response of a single molecule carries directly to dense monolayers, suggesting that suitably engineered films could translate the single-molecule advantage into device-level performance. Taken together, our results identify nodal density as a central design principle for engineering quantum-interference-enhanced thermoelectric materials based on series-connected molecular architectures.

\vspace{6pt}

\funding{This research was graciously supported by the National Science Foundation (NSF) under award number QIS-2412920.}

\institutionalreview{Not applicable.}

\informedconsent{Not applicable.}

\dataavailability{The original contributions presented in the study are included in the article; further inquiries can be directed to the corresponding author.}

\conflictsofinterest{The author declares no conflicts of interest. The funders had no role in the design of the study; in the collection, analyses, or interpretation of data; in the writing of the manuscript; or in the decision to publish the results.}

\appendixtitles{yes} 
\appendixstart
\appendix

\section{Analytic Solutions for the Low-Energy Split-Node Model}\label{appendixa}

We model the low-energy \(\pi\)-channel by an equispaced set of quadratic nodes,
\begin{equation}
\label{eq:SI_Tpoly}
\mathcal{T}_\pi(E)\propto E^2 \prod_{m=1}^{M} (E-m\Delta)^2 (E+m\Delta)^2,
\end{equation}
which represents \(N=2M~{+}~1\) nodes centered at \(E=0\) with spacing \(\Delta\).  
The Onsager functions
\begin{equation}
\label{eq:SI_Lnu_def}
\Lfn{\nu}=\frac{1}{h}\int (E-\mu)^{\nu}\,\mathcal{T}_\pi(E)\,
\Bigl(-\frac{\partial f_0}{\partial E}\Bigr)\,dE,
\end{equation}
with  $f_0(E) = \left[\exp((E-\mu_0)/kT_0) + 1 \right]^{-1}$,  admit closed forms because
\begin{equation}
\label{eq:SI_dfdE}
-\frac{\partial f_0}{\partial E}=\frac{1}{4k_B T_0}\,\mathrm{sech}^2\!\left(\frac{E-\mu_0}{2k_B T_0}\right)
\end{equation}
has finite even central moments.
For the minimal three-node case (\(N=3\), \(M=1\)),
\begin{equation}
\label{eq:SI_Tpoly_N3}
\mathcal{T}_\pi(E)\propto (E-\Delta)^2 E^2 (E+\Delta)^2
=E^6-2\Delta^2 E^4+\Delta^4 E^2.
\end{equation}
Writing \(\beta=(k_B T_0)^{-1}\) and factoring a scale \(\mathcal{N}\),
\begin{adjustwidth}{-\extralength}{0cm}
\begin{align}
\label{eq:SI_L0}
\frac{\Lfn{0}}{\mathcal{N}}&=
\mu^6-2\Delta^2\mu^4+\Delta^4\mu^2
+\frac{\pi^2}{3}\beta^{-2}\!\bigl(\Delta^4-12\Delta^2\mu^2+15\mu^4\bigr)
+\frac{7\pi^4}{15}\beta^{-4}\!\bigl(15\mu^2-2\Delta^2\bigr)
+\frac{31\pi^6}{21}\beta^{-6}, \\
\label{eq:SI_L1}
\frac{\Lfn{1}}{\mathcal{N}}&=
\mu\Biggl[
\frac{2\pi^2}{3}\beta^{-2}\bigl(\Delta^4-4\Delta^2\mu^2+3\mu^4\bigr)
+\frac{28\pi^4}{15}\beta^{-4}\bigl(5\mu^2-2\Delta^2\bigr)
+\frac{62\pi^6}{7}\beta^{-6}
\Biggr].
\end{align}
\end{adjustwidth}
The thermopower then follows as
\begin{equation}
\label{eq:SI_S}
S(\mu,\Delta,T)= -\frac{1}{eT}\,\frac{\Lfn{1}}{\Lfn{0}}.
\end{equation}

To locate the thermopower optimum, we solve \(\partial_\mu S=0\). Approximating the peak position by the single-quadratic-node result,
\begin{equation}
    \label{eq:SI_single_node_peak}
    |\mu_{\rm peak}-\mu_{\rm node}|=\frac{\pi}{\sqrt{3}}\,k_B T_0,
\end{equation}
yields, for \(N=3\),
\begin{equation}
\label{eq:SI_Delta_peak_S}
\Delta_{\rm peak}^{\rm S}=\frac{\pi}{3\beta}\sqrt{\frac{230-26\sqrt{37}}{7}}
\;\approx\;3.36\,k_B T_0,
\end{equation}
\begin{equation}
    \label{eq:SI_mu_peak_S}
    \mu_{\rm peak}^{S}\approx 1.92\,k_B T_0,
\end{equation}
Direct numerical maximization of \eqref{eq:SI_S} confirms that \(\Delta_{\rm peak}^{\rm S}/k_B T_0\simeq 3.4\) is nearly independent of \(N\): additional equispaced nodes chiefly increase peak magnitude and broaden the response, while the optimal spacing is set locally by the nearest neighbors at \(\pm\Delta\).

For the electronic figure of merit,
\begin{equation}
\label{eq:SI_ZTel_def}
ZT_{\rm el}=\frac{S^2 G T_0}{\kappa_e},\qquad
\kappa_e=\frac{1}{T_0}\left(\Lfn{2}-\frac{\left[\Lfn{1}\right]^2}{\Lfn{0}}\right),
\end{equation}
the same method yields
\begin{equation}
        \Delta_{\rm peak}^{\rm ZT}\approx 3.52\,k_B T_0,\qquad
        \mu_{\rm peak}^{\rm ZT}\approx 3.14\,k_B T_0,
\end{equation}
The optimum spacing is slightly larger for $ZT_{\rm e}$ than for $S$, but both lie in the narrow window $\Delta\simeq 3.4$–$3.5\,k_B T_0$.

\subsection*{Effect of a $\sigma$ Background}

A weak, energy-flat $\sigma$ channel enters additively,
\begin{equation}
\label{eq:SI_T_total_sigma}
\mathcal{T}_{\rm tot}(E)=\mathcal{T}_\pi(E)+\mathcal{T}_\sigma,
\end{equation}
modifying the Onsager functions as
\begin{equation}
\label{eq:SI_L_sigma_split}
\Lfn{0}\to \Lfn{0}_\pi+\mathcal{T}_\sigma,\qquad 
\Lfn{1}\to \Lfn{1}_\pi,
\end{equation}
and hence the thermopower becomes
\begin{equation}
\label{eq:SI_S_sigma}
S_{\rm tot}(\mu,\Delta,T)=\frac{S_\pi(\mu,\Delta,T)}{1+\rho},\qquad
\rho=\frac{\mathcal{T}_\sigma}{\Lfn{0}_\pi(\mu^*,\Delta^*,T)}.
\end{equation}
Thus, the $\sigma$ background enters only through \(\Lfn{0}\), acting as a parallel conductance that dilutes the $\pi$-system signal. The suppression is linear in $S$, but because $ZT_{\rm e}\propto S^2$, the degradation in thermoelectric performance is quadratic: even modest $\rho$ reduce $ZT_{\rm e}$ disproportionately.

Crucially, $\Delta_{\rm peak}$ remains unchanged to leading order. This robustness reflects the fact that \(\Lfn{1}\), which sets the numerator of $S$, depends on the odd part of the transmission and is unaffected by a flat background, whereas \(\Lfn{0}\) and \(\Lfn{2}\) acquire additive $\sigma$ terms that reduce the magnitude but not the location of the optimum. Physically, leakage through the $\sigma$ system sets the ceiling on the achievable enhancement but does not obscure where interference-driven amplification occurs. Suppressing $\mathcal{T}_\sigma$ by one to two orders of magnitude is therefore sufficient for the nodal response to remain visible, consistent with the numerical results of Figures~\ref{fig:NCC} and~\ref{fig:NCCA}.

\bibliography{refs_sent_by_entropy}

\end{document}